\documentstyle[12pt]{article}
\font\srm=cmr9

\def\nabl{\nabla\!}
\def\beq{\begin{equation}} \def\eeq{\end{equation}} \def\eqn{\label}
  \def\M{{\cal M}} 
\def\A{{\cal A}}\def\AL{{\cal A}^{_{\rm L}}} \def\AM{{\cal A}^{_{\rm M}}}
\def\H{{\cal H}} \def\HL{{\cal H}^{_{\rm L}}} \def\HM{{\cal H}^{_{\rm M}}} 
\def\D{{D}} \def\DL{{D}^{_{\rm L}}}  

\def\j{J}
\def\fX{f^{_{\rm X}}} \def\wX{w^{_{\rm X}}}
\def\nX{n_{_{\rm X}}} 
\def\bn{b_{\rm n}} \def\bp{b_{\rm p}}
\def\bX{b_{_{\rm X}}}
\def\muX{\mu^{_{\rm X}}} 
 \def\mp{m_{\rm p}} \def\me{m_{\rm e}}

\def\nn{n_{\rm n}} \def\np{n_{\rm p}} \def\ne{n_{\rm e}}
\def\en{e^{\rm n}} \def\ep{e^{\rm p}} \def\ee{e^{\rm e}}
\def\eX{e^{_{\rm X}}} 
\def\chiX{\chi_{_{\rm X}}}
\def\pin{\pi^{\rm n}} \def\pip{\pi^{\rm p}} 
\def\piX{\pi^{_{\rm X}}} 
\def\wn{w^{\rm n}} \def\wp{w^{\rm p}} \def\we{w^{\rm e}}
\def\fn{f^{\rm n}} \def\fp{f^{\rm p}} \def\fe{f^{\rm e}}

\def\alphX{\theta_{_{\rm X}}}
  
\def\xiX{\xi_{_{\rm X}}}  \def\lamX{\lambda_{_{\rm X}}}
\def\lamn{\lambda_{\rm n}} \def\lamp{\lambda_{\rm p}}
\def\lame{\lambda_{_{\rm e}}}      
\def\mun{\mu^{\rm n}} \def\mup{\mu^{\rm p}} \def\mue{\mu^{\rm e}}
\def\phin{\varphi^{\rm n}} \def\phip{\varphi^{\rm p}}
\def\alpn{\alpha^{\rm p}_{\rm n}}
\def\muL{\mu^{_{\rm L}}\,\!} \def\lL{\ell}
\def\omeg{\Omega}

\begin{document}

\title
{Superconducting Superfluids in Neutron Stars}

\author { {\bf Brandon Carter}
\\ \hskip 1 cm\\
D\'epartement d'Astrophysique Relativiste et de Cosmologie,\\
Centre National de la Recherche Scientifique, \\Observatoire de
Paris, 92195 Meudon, France.}

\date{\it March 2000} 

\maketitle 

{\bf Abstract:}
For treatment of the layers below the crust of a neutron star it is useful 
to employ a relativistic model involving three independently moving 
constituents, representing superfluid neutrons, superfluid protons, and 
degenerate negatively charged leptons. A Kalb Ramond type formulation
is used here to develop such a model for the specific purpose of 
application at the semi macroscopic level characterised by lengthscales 
that are long compared with the separation between the highly localised 
and densely packed proton vortices of the Abrikosov type lattice that 
carries the main part of the magnetic flux, but that are short compared 
with the separation between the neutron vortices.

\vskip 2 cm

\section{Introduction}
\label{Sec1}

The purpose of this article is to present a concise overview of a class
of macroscopic relativistic superconducting superfluid models
developed\cite{C85,CL98}  as a generalisation of previous non
conducting relativistic superfluid models\cite{LK82,CK92b,CL95c} with a
view to applications concerning the layers below the crust of a neutron
star, which are believed to be well described by three constituent
superconducting superfluid models of the kind that was introduced (as a
charged generalisation of the Andreev - Bashkin
model\cite{AndreevBashkin76})for a superfluid mixture) by Vardanyan and
Sedrakyan\cite{VarSed81}, and that has more recently been further
developed (though still using a non-relativistic treatment) by Mendell
and Lindblom\cite{MenLin91}.  The three basic ingredients in a
description of this kind are, firstly, a condensate of superfluid
neutrons, secondly an independently moving -- effectively superconducting 
-- condensate  of superfluid protons, and thirdly a negatively charged
degenerate leptonic constituent (consisting mainly of electrons but
also including a significant proportion of muons) that is of
``normal'', i.e. non-superfluid, kind. Such a treatment does not
include thermal effects (whose inclusion would involve a fourth
constituent representing entropy) but should nevertheless be applicable
as a very good first approximation except during a short lived  high
temperature phase immediately after the birth of the neutron star.

As the relativistic analogue of the kind of phenomenological description
introduced in a Newtonian context by Bekarevich and Khalatnikov\cite{BekaK61}
(as a generalisation of Landau's original two-constituent model) it will 
first be shown how to set up a general category of three-constituent 
perfectly conducting perfect fluid models of the type that is needed, as a 
preliminary for the more specific developments that follow. This 
category\cite{CL98} includes, as a specialisation, the case in which the 
neutronic and the protonic constituents are both characterised by strictly 
irrotational behaviour of the kind that is relevant in neutron stars on a 
``mesoscopic'' scale, meaning a scale large compared  with that of the 
underlying microscopic particle description, but small compared with the 
macroscopic scale separation between the vortex defects within which the 
superfluid comportment is locally violated. Models of this irrotational 
kind are just a specially simple limit within the more general category 
that is needed for the purpose of treating the superconducting superfluid 
on a ``macroscopic'' scale, meaning a scale that is large compared with 
the separation between vortices. 

The present discussion will be focussed on an intermediate ``semi
macroscopic'' scale meaning a scale that is small compared with the
spacing between the  superfluid neutron vortices (which will be rather
widely separated due to the relatively low angular velocity of the
star, though they contain quite a lot of energy due to their ``global''
nature) but large compared with the Abrikosov lattice spacing between
the ``local'' proton vortices, which are expected to be much more
numerous in order to carry the rather large magnetic fluxes that are
thought to be present.

\section{Generic category of 3-constituent superconducting superfluid models}
\label{Sec2}

In so far as its contribution to the mass density is concerned, the
most important of the the three independent constituents under
consideration is that of the superfluid neutrons, with number current
4-vector $\nn^{\,\rho}$, say. The second contribution is that of the
superconducting protons -- which make up a small but significant part
of the mass density -- with number current four-vector $\np^{\,\rho}$.
The third constituent is that of the degenerate non-superconducting
background of negatively charged leptons -- consisting mainly of
electrons, but including also a certain proportion of muons at the high
densities under consideration -- with a corresponding lepton number
number current vector $\ne^{\,\rho}$ say.  This negatively charged
``normal'' (i.e. non superconducting) constituent contributes only a
very small fraction of the mass density, but it nevertheless has a
crucially important role, not just because the corresponding unit
vector $u^\rho$ defined by
 \beq \ne^\rho=\ne u^\rho\, ,\hskip 1 cm \ne^2=-\ne^\rho{\ne}_\rho
 \eqn{02}\eeq
characterises the natural reference frame of rigid corotation in an
equilibrium configuration, but more generally, in so far as electromagnetic 
effects are concerned, because in terms of the electron charge coupling 
constant $e$ the corresponding total electric current 4-vector 
will be given by 
 \beq \j^\mu=e\big(\np^{\,\rho}-\ne^{\,\rho}\big) \, .\eqn{1}\eeq

It will be convenient to express formulae such as this in a condensed
notation system based on the use of the summation convention for
``chemical'' indices represented by capital Latin letters running over
the three relevant values, namely {\srm X}=n, {\srm X}=p, {\srm X}=e.
Using this convention, the equation (\ref{1}) for the electric current
density can be rewritten in the consise form
 \beq \j^\rho=\eX \nX^{\,\rho} \, ,
 \eqn{8}\eeq where the the charges per neutron, proton, and electron
are given respectively by  $\en=0$, $\ep=e$, and $\ee=-e$.

Since each of the three independent currents involved is conserved, it
will be possible to use a Kalb-Ramond formulation in which, instead of
imposing the three corresponding conservation laws 
  \beq \nabl_{\!\rho}\, \nX^{\,\rho}=0\, ,  \eqn{00}\eeq 
as dynamical equations, they will be obtained as
identities by postulating that the currents should have the form
 \beq \nX^{\,\rho}=\nabl_\sigma\, \bX^{\,\rho\sigma}\, ,
 \eqn{70}\eeq 
for corresponding antisymmetric gauge bivector fields  
$\bX^{\,\rho\sigma}$ which are physically defined only modulo Kalb
Ramond gauge transformations of the  form
 \beq \bX{^{\rho\sigma}} \, \mapsto \, \bX{^{\rho\sigma}}+
 \nabl_{\!\nu}\,\alphX{^{\nu\rho\sigma}}\, . \eqn{69}\eeq 
for arbitrary antisymmetric trivector fields
$\alphX{^{\nu\rho\sigma}}$.

Since our present treatment will be restricted to the conservative limit in
which dissipative effects are neglected, the analysis will be assumed to be
expressible in terms of a variational principle based on a Lagrangian
density, in which as usual the representation of the electromagnetic field 
requires the introduction of a Maxwellian gauge 1-form $A_\rho$, in terms 
of which the gauge invariant electromagnetic field tensor is given by
 \beq F_{\rho\sigma}=2\nabl_{[\rho}A_{\sigma]}\, , \eqn{11}\eeq
(using square brackets to indicate index antisymetrisation). The
implementation of the Kalb Ramond formulation requires that the set of
independent currents   $\nX^{\,\rho}$  be supplemented\cite{CL98} by the 
introduction of a corresponding set of vorticity 2-forms 
$\wX{_{\rho\sigma}}$ each of which is characterised both by an algebraic 
degeneracy condition of the form
 \beq w_{[\mu\nu}w_{\rho]\sigma}=0\, , \eqn{60}\eeq
and by a  closure condition of the form
 \beq \nabl_{[\nu}\,\wX_{\rho\sigma]}=0\,, \eqn{61}\eeq
so that each such 2-form $w_{\rho\sigma}$ is interpretable as a pullback of 
a prescribed area measure on a two-dimensional base space. This means that 
in terms of suitably chosen local vorticity base space cordinates 
$\chi^{_1}$, $\chi^{_2}$, the corresponding pair of scalar fields 
$\chiX^{_1}$,  $\chiX^{_2}$ induced on the four dimensional spacetime 
background will specify the vorticities by prescriptions of the form
$w_{\rho\sigma}=2\chi^1_{,[\rho}\chi^2_{,\sigma]}$.

In terms of these quantities, a Lagrangian of the appropriate kind will
will be expressible in the generic form
 \beq {\cal L}=\Lambda+\j^\rho A_\rho +{_1\over^2}\bX^{\,\sigma\rho}
 \wX_{\,\rho\sigma}\, ,\eqn{74}\eeq which consists of a pair of gauge
dependent coupling terms preceded by a first term $\Lambda$ that is a
function only of the relevant gauge independent field quantities, which
in addition, of course to the spacetime metric $g_{\rho\sigma}$ and the
electromagnetic field tensor $F_{\rho\sigma}$ consist of the three
independent currents $ \nX^{\,\rho}$ and the three corresponding
vorticity 2- forms  $\wX_{\,\rho\sigma}$. This means that its most
general infinitesimal variation will be given by an expression of the
form
 \beq \delta\Lambda=\muX_{\,\rho}\,\delta \nX^{\,\rho} +{_1\over^2}
\lamX^{\sigma\rho}\,\delta\wX_{\rho\sigma} +{1\over
8\pi}\H^{\sigma\rho}\,\delta F_{\rho\sigma} +{\partial
\Lambda\over\partial g_{\rho\sigma}}\,\delta g_{\rho\sigma}\, ,
\eqn{75}\eeq where the coefficients of the metric variations are not
independent of the others but must satisfy a Noether type identity of
the form
 \beq {\partial \Lambda\over\partial g_{\rho\sigma}}= {\partial\Lambda
 \over\partial g_{\sigma\rho}}= {_1\over^2}\muX_{\,\nu} \nX^{\,\rho}
 g^{\nu\sigma}+{_1\over^2}\lamX^{\,\nu\rho}\wX_{\,\nu}{^\sigma} 
 +{1\over 16\pi}\H^{\nu\rho} F_\nu{^\sigma}\, . \eqn{76}\eeq 
The triplet of covectorial quantities  $\muX_{\,\rho}$ is simply
interpretable as representing usual 4-momenta (per particle) of the
neutrons, protons, and leptons.  The triplet of rather less familiar
bivectorial coefficients $\lamX^{\,\sigma\rho} = -\lamX^{\,\rho\sigma}$
in this expansion characterises the macroscopic anisotropy arising
respectively from the concentration of energy and tension in
mesoscopic  vortices of the neutron and proton superfluids, as a
consequence of their vorticity quantisation conditions, in the manner
discussed in our previous work\cite{CL95c} on the single constituent
model.  These new four dimensional bivectorial coefficients replace the
three dimensional (space) vectorial coefficients introduced for a
similar purpose in a more restricted Newtonian framework by Bekarevich
and Khalatnikov\cite{BekaK61}.  Finally the bivectorial coefficient
$\H^{\rho\sigma}= -\H^{\rho\sigma}$ will be interpretable as an
electromagnetic displacement tensor, in terms of which the total
electromagnetic field tensor (\ref{11}) will be given by an expression
of the form
 \beq F^{\rho\sigma}= \H^{\rho\sigma} + 4\pi \M^{\rho\sigma}\, , 
 \eqn{77}\eeq 
in which $\M^{\rho\sigma}$ is what can be interpreted as the magnetic
polarisation tensor. In the application that we are considering, this --
typically dominant -- Abrikosov polarisation contribution 
$4\pi\M^{\rho\sigma}$ is to be thought of as representing the part of the 
magnetic field confined in the vortices, while the -- typically much 
smaller -- remainder $\H^{\rho\sigma}$ represents the average contribution 
from the  field in between the vortices, which can be expected to vanish
by the ``Meissner effect'' in strictly static configurations, but which
can be expected to acquire a non zero value  in rotating configurations due
to the London mechanism that will be discussed below.

In the application of such a generalised Kalb Ramond type variation
principle, the gauge fields  $\bX^{\,\sigma\rho}$ and $A_\rho$ are to 
be considered as free variables, but $ \nX^{\,\rho}$ and
$\wX_{\,\rho\sigma}$ are not. Each current $ \nX^{\,\rho}$ is to be
considered as fully determined by the corresponding gauge bivector
$\bX^{\,\rho\sigma}$ according to the prescription (\ref{70}) while
each vorticity 2 form $\wX_{\,\rho\sigma}$ is to be considered as
being determined by corresponding freely chosen scalar base coordinate
pullback fields $\chiX^{_1}$, $\chiX^{_2}$, which means that the
variation of any vorticity 2 form $w_{\rho\sigma}$ will be determined
by a corresponding freely chosen displacement vector field
$\xi^\rho$ according to a prescription\cite{CL95c} of the form
$\delta w_{\mu\nu}=-2\nabl_{[\mu}\big( w_{\nu]\rho} \xi^\rho\big)$.

Subject to these rules, the ``diamond'' variational integrand
 \beq \diamondsuit {\cal L}=\Vert g \Vert^{-1/2}\, \delta\Big(\Vert g 
 \Vert^{1/2}{\cal L} \Big) = \delta{\cal L} +{_1\over^2} {\cal L} 
 g^{\mu\nu}\, \delta g_{\mu\nu}\, . \eqn{27}\eeq
needed for the application of the variational principle will be given by
 \beq \diamondsuit {\cal L}=\big(\nabl_\rho\,\piX_{\,\sigma}- {_1\over^2}
 \wX_{\,\rho\sigma}\big)\delta \bX^{\,\rho\sigma}+ \fX_{\,\rho}\, 
 \xiX^{\,\rho}+\Big(\j^\rho-{1\over 4\pi}\nabl_\sigma\, \H^{\rho\sigma}
 \Big)\,\delta A_\rho +{_1\over^2} T^{\mu\nu}\,\delta g_{\mu\nu} 
 +\nabl_\sigma\, {\cal R}^{\ \sigma} \, , \eqn{80}\eeq
in which it is useful to allow for the possibility of varying the
background spacetime metric $g_{\rho\sigma}$, not only for the purpose 
of dealing with cases in which one may be concerned with General 
Relativistic gravitational coupling, but even for dealing with cases in 
which one is concerned only with a flat Minkowski background, since, as 
will be made explicit below, the effect of virtual virtuations with respect 
to the relevant curved or flat background can be used for evaluating the 
relevant ``geometric'' stress energy momentum density tensor 
$T^{\rho\sigma}$. The coefficients of the current variations are the 
usual gauge dependent total momentum covectors given by
 \beq \piX{_{\!\rho}} =  \muX{_{\!\rho}}+\eX A_\rho\, .\eqn{14}\eeq
The coefficients $\fX_{\,\rho}$ of the three independent displacement
displacement vector fields $\xiX^{\,\rho}$ will be interpretable as
the effective force densities acting on the corresponding constituent 
currents. The generic expression for these force densities can be 
read out for the neutrons {\srm X}=n, protons  and {\srm X}=p and
leptons {\srm X}=e respectively as
 \beq \fn_{\rho} =\big(\nn^{\,\sigma}+\nabl_\nu\,
 \lamn^{\,\sigma\nu}\big) \wn_{\, \sigma\rho}\, ,\eqn{81n}\eeq
 \beq    \fp_{\rho} = \big(\np^{\,\sigma} +\nabl_\nu\,
 \lamp^{\,\sigma\nu}\big)\wp_{\, \sigma\rho}\, ,\eqn{81p}\eeq
 \beq   \fe_{\rho} = \big(\ne^{\,\sigma} +\nabl_\nu\,
 \lame^{\,\sigma\nu}\big)\we_{\, \sigma\rho}\, . \eqn{81e}\eeq 
Although it is of no relevance for the application of the variation 
principle, it can be noted for the record that the current appearing
in the final divergence term of (\ref{80}) will be given by 
 \beq {\cal R}^{\ \sigma}=\piX_{\,\rho}\,\delta\bX^{\,\rho\sigma}
 -\big(\bX^{\,\rho\sigma}+\lamX^{\,\rho\sigma}\big)\wX_{\,\rho\nu}
 \xiX^\nu +{_1\over^2}\piX_{\,\rho}\bX^{\,\rho\sigma} g^{\mu\nu}\, \delta 
 g_{\mu\nu}+{1\over 4\pi}H^{\sigma\nu}\,\delta A_\nu\, .  \eqn{82}\eeq
An entity of much greater practical interest is the corresponding stress
momentum energy density tensor, which can be seen to be given by
 \beq T_{\sigma}^{\ \rho}=\nX^{\,\rho}\muX_{\,\sigma}+\lamX^{\,\nu\rho}
 \wX_{\,\nu\sigma} + {1\over 8\pi}\H^{\nu\rho} F_{\nu\sigma}+\Psi 
 g^\rho_{\ \sigma}\, , \eqn{83}\eeq
where the generalised pressure function is given by
 \beq \Psi=\Lambda-\nX^{\,\nu}\muX_{\,\nu}+\bX^{\,\rho\sigma} \big(
 \nabl_{\,\rho}\,\piX_{\,\sigma}-{_1\over^2}\wX_{\,\rho\sigma}\big)
 \, .\eqn{84}\eeq
The last term in (\ref{80})  will evidently drop out when we impose the
condition of invariance with respect to infinitesimal variations of the
bivectorial gauge potentials $\bn^{\,\rho\sigma}$ and $\bp^{\,\rho\sigma}$ 
is imposed, a requirement which can be seen from (\ref{80}) to give field
equations of the form
 \beq \wX_{\,\rho\sigma}=2\nabl_{[\rho}\piX{_{\!\sigma]}}
 =2\nabl_{[\rho}\muX{_{\!\sigma]}}+\eX F_{\rho\sigma}\, , \eqn{85} \eeq
which are evidently equivalent to what in other formulations could be
considered just as defining relations for the vorticity two-forms. 
The remaining field equations obtained from (\ref{80}) will consist of
 \beq \nabl_\sigma\, \H^{\rho\sigma}=4\pi \j^\rho\, , \eqn{86} \eeq
together with the condition that the force density coefficients should all
vanish, i.e.
 \beq \fX_{\,\rho}=0 \eqn{39}\eeq 
for each of the three relevant chemical index values {\srm X}=n, 
{\srm X}=p, {\srm X}=e. 

\section{The semi-macroscopic application.}
\label{Sec6}

To be more specific, we need to specify the scale of application for
which the model is intended. At a mesoscopic level, meaning on scales
large compared with the dimensions of individual molecules or Cooper type
pairs, but small compared with the intervortex spacing, the 
appropriately specialised model will be of purely fluid type, meaning
that the function $\Lambda$ should not depend on the vorticity forms
$\wX_{\,\rho\sigma}$ which implies the vanishing of the
Bekarevich - Khalatnikov coefficients, i.e. the restriction
$\lamX^{\,\sigma\rho}=0  $.
A further restriction that applies at this mesoscopic level is that
for the superfluid constituents, namely the neutrons and the protons,
but not for the degenerate lepton constituent, the corresponding
vorticities themselves should be zero, i.e. for
{\srm X} $\neq {\rm e}$ we should have $ \wX_{\,\rho\sigma}=0$,
which is the integrability condition for the corresponding momenta
to have the form 
$ 2\pin_{\,\rho}={\hbar}\nabl_\rho\,\phin $, 
$ 2\pip_{\,\rho}={\hbar}\nabl_\rho\,\phip$
in which the scalars $\phin$ and $\phip$ will be interpretable as the
phases of underlying bosonic quantum condensates, with periodicity
$2\pi$, and in which the preceeding factors of 2 have been inserted
to allow for the fact that the relevant bosons will consist not of single
neutrons and protons but of Cooper type pairs thereof.

At a much larger, fully macroscopic scale, involving averaging over
large numbers of the neutron and proton vortices (that arise as topological
defects of the mesoscopic phase fields) the neutronic and protonic
constituents will be characterised by not just by non vanishing 
effective large scale vorticities $\wn_{\,\rho\sigma}\neq 0$ and 
$\wp_{\,\rho\sigma}\neq 0$,  but also by non vanishing Bekarevich 
- Khalatnikov coefficients, $\lamn^{\,\sigma\rho}\neq0  $ and  
$\lamp^{\,\sigma\rho}\neq0  $, so that only the degenerate electrons
still behave in an effectively fluid manner, in accordance with
the restriction
\beq \lame^{\,\sigma\rho}= 0 \, .\eqn{026e} \eeq

The purpose of the present article is to focus on an intermediate scale
that will be refferred to as ``semi macroscopic'' meaning that it deals
with averages over scales that are large compared with the spacing
between proton vortices, but small compared with the spacing between
the neutron vortices, which are expected to be relatively widely spaced
in typical circumstances within neutron stars, whose angular velocities
are very low as measured by local physical timescales, whereas their
magnetic fields are typically rather large. On such  ``semi
macroscopic'' scales the behaviour of the neutrons constituent will not
just be of strictly fluid type, meaning that it will be characterised
by
 \beq \lamn^{\,\sigma\rho}= 0 \, ,\eqn{026n} \eeq but it will also be
subject to the mesoscopic superfluidity condition
 \beq 2\pin_{\,\rho}={\hbar}\nabl_\rho\,\phin\, , \eqn{47}\eeq and
hence 
 \beq \wn_{\,\rho\sigma}=  0\, ,\eqn{47b}\eeq 
so that the corresponding dynamical equation, i.e. the requirement
(\ref{39}) that the effect that the net force density (\ref{81n})
should vanish, will be automatically satisfied everywhere outside the
microscopic cores of the neutron superfluid votices (which are of
``global'' type, meaning that their energy would diverge
logarithmically in the absence of the ``infra red'' cut off imposed by
the presence of neighbouring vortices). However since we are
considering scales large compared with the separation distance between
the much more numerous proton vortices (which are of ``local'' type,
meaning that their energy density falls off exponentially on a
microscopic lengthscale $\lL$ whose evaluation will be discussed below)
the proton constituent will be characterised not only by non vanishing
averaged vorticity,  $\wp_{\,\rho\sigma}\neq 0$ but also by a non
vanishing Bekarevich - Khalatnikov coefficient, 
$\lamp^{\,\sigma\rho}\neq 0$.  This means that the dynamical
requirement that the corresponding net force density (\ref{81p}) should
vanish will provide a rather complicated dynamical equation,
expressible in the form
 \beq 2\np^{\,\sigma}\nabl_{[\sigma}\,\mup_{\rho]}+e\np^{\,\sigma} 
 F_{\sigma\rho} +\wp_{\,\sigma\rho}\nabl_\nu\,\lamp^{\,\sigma\nu}=0  
 \, , \eqn{90}\eeq
in which the first term is interpretable as the negative of the Joukovski
force density due to the ``Magnus effect'' acting on the proton vortices,
the middle term is the Lorentz force density representing the effect of the 
magnetic field on the passing protons, while the last term (which was 
absent in the mesoscopic description) represents the extra force density 
on the fluid due to the effect of the tension of the vortices. As a 
consequence of its ``normality'' property (\ref{026e}), the leptonic 
constituent is governed by a dynamical equation of the simpler form
\beq
2\ne^{\,\sigma}\nabl_{[\sigma}\,\mue_{\rho]}+e\ne^{\,\sigma} F_{\sigma\rho}
=0  \, .  \eqn{90e}\eeq

In view of the highly localised nature of the proton vortices,
it is reasonable\cite{CPL99} to suppose that their action contribution 
should be fully determined just by the Abrikosov lattice density of 
such vortices, and that the only independent contribution from the
electromagnetic field $F_{\rho\sigma}$ should be the part provided
by the residual -- weaker but much more widely extended -- part
of the flux outside the vortex tubes, as given by the field
$\H^{\rho\sigma}$ that is given by the relation (\ref{77}), on the 
understanding that the polarisation contribution
$4\pi \M^{\rho\sigma}$ represents the part of the flux attributable
to the the vortex tubes. This implies that that the gauge independent
term $\Lambda$ in the action will be decomposable in the form
 \beq \Lambda=\Lambda_{\rm MV}+\Lambda_{\rm F}\, ,\eqn{160}\eeq
where the macroscopic contribution $\Lambda_{\rm MV}$ is required to be
functionally independent of $F_{\mu\nu}$, so that it is determined just 
by the vorticity $\wp_{\,\rho\sigma}$ and the currents $\nX^{\,\rho}$, 
while for constency with the variational definition (\ref{75}) there
will be no  loss of generality in taking the remaining, electromagnetic
field dependent, contribution to have the simple quadratic form
 \beq \Lambda_{\rm F}={1\over 16\pi} \H_{\rho\sigma}\H^{\sigma\rho}\, .
 \eqn{161}\eeq
which, in the absence of the polarisation contribution
$4\pi \M^{\rho\sigma}$ in (\ref{77}), would reduce just to the
usual action contribution for an electromagnetic field in vacuum.

Despite of being much more densely packed than the neutron vortices, 
the fact that (unlike the neutron vortices) the proton vortices are 
exponentially localised within a microscopic confinement radius 
means that their mutual interactions (unlike those of the long 
range interacting neutron vortices) should remain entirely negligible 
even for extremely high magnetic fields, so that their contribution 
to the action should be simply proportional to their density. This 
means that it will be possible to make the further decomposition
 \beq \Lambda_{\rm MV}=\Lambda_{\rm M}+\Lambda_{\rm V}\, , \eqn{162}\eeq
in which $\Lambda_{\rm M}$ is entirely independent of the vorticity 
$\wp_{\,\rho\sigma}$, so that it depends only on the
three independent currents $\nX^{\,\rho}$, while the remainder 
$\Lambda_{\rm V}$ is just linearly proportional to the protonic 
vorticity density, so that it will be expressible in the form
 \beq\Lambda_{\rm V}=\lamp\wp\eqn{200}\eeq
where $\wp$ is the protonic vorticity magnitude as defined by 
 \beq \wp=\sqrt{{\wp}{^{\,\rho\sigma}} \wp{_{\rho\sigma}}/2}
 \, ,\eqn{201} \eeq
and where, like $\Lambda_{\rm M}$, the coefficient $\lamp$
depends only on the currents $\nX^{\,\rho}$. On the basis of
dimensional considerations  -- which should be valid provided the
Ginzburg landau ratio of the  London penetration length $\lL$ that 
will be discussed below to the relevant Pippard correlation length 
is not too far from the order of unity value that characterises the 
Bogomol'nyi limit\cite{CPL99b} -- it can be anticipated that, as is 
confirmed by more detailed analysis\cite{Men91,Men98,CPL99} the 
Bekarevich - Khalatnikov coefficient $\lamp$ will have an order of 
magnitude given in terms of the relevant charged particle mass, which 
in the neutron star case under consideration is the proton mass $\mp$ 
(but which in an ordinary metallic superconductor would be the 
electron mass $\me$) by by $\lambda\approx \hbar \np/\mp$.

The appropriate form for the polarisation tensor
$4\pi \M^{\rho\sigma}$ in (\ref{77}), can be seen by decomposing
the vector potential $A_\rho$ as the sum of a gauge dependent
contribution proportional to the proton momentum covector $\pip_\rho$ 
and a gauge independent remainder ${\cal A}_\rho$ in the form 
 \beq A_\rho= {1\over e}\pip_{\,\rho}+\A_\rho\,  ,\eqn{202}\eeq
from which, by extrior differentiation, one obtains a corresponding
decomposition of the form
 \beq F_{\rho\sigma}={1\over e}\wp_{\rho\sigma}+\H_{\rho\sigma}
 \eqn{203}\, ,\eeq
with
 \beq \H_{\rho\sigma}=2\nabla_{\![\rho}\A_{\sigma]} \eqn{204}\, .\eeq
This decomposition has the required form  (\ref{77}) provided one makes 
the identification
 \beq \M_{\rho\sigma}={1\over 4\pi e}\wp_{\rho\sigma}\, ,\eqn{205}\eeq
for what will be referred to as the Abrikosov polarisation tensor.

\section{Phenomenological interpretation.}
\label{Section 7}

As discussed in more detail in particular cases\cite{CL98,CPL99}
(correcting earlier work\cite{Men91,Men98}
in which it was overhastily assumed to cancel out)
the quantity $4\pi\M_{\rho\sigma}$ defined by (\ref{205}) will be 
interpretable as representing the part of the magnetic flux confined 
to the proton vortices, whose action contribution will be included in 
the term $\Lambda_{\rm V}$ given by (\ref{200}), while 
$\H_{\rho\sigma}$ accounts for the remainder of the flux, which will 
be distributed over the region outside  the proton vortices, and 
whose contribution to the action will be given by (\ref{161}). The 
covector $\A$ will be given by
 \beq \A_\rho=-{1\over e}\mup_\rho\, ,\eqn{206}\eeq
so that it will be obtainable from the equation of state function 
for $\mup_\rho$ as derived from  $\Lambda_{\rm M}$ as a linear 
combination of the form
 \beq \A_\rho=\A^{\rm n}_{\,\rho}+\AL_{\,\rho}+\AM_{\,\rho}
 \, ,\eqn{207}\eeq
in which the terms are proportional respectively to the neutron
4-momentum,  the ``normal'' reference state unit vector $u^\rho$ 
(as specified according to (\ref{02}) by the leptonic current), 
and the (semi macroscopic) electric current $\j^\rho$, so that they 
will be expressible as
 \beq  \A^{\rm n}_{\,\rho}={1\over e}\alpn \,\pin_\rho\, ,
 \hskip 1cm \AL_{\,\rho} =-{\muL\over e}\, u_\rho\, ,
 \hskip 1cm \AM_{\,\rho}= -4\pi\lL^2 \j_\rho \, , \eqn{210}\eeq
with proportionality factors that depend (just) on the form of the master
function specfying $\Lambda_{\rm M}$ in terms of the relevant currents.
In particular the dimensionless parameter $\alpha$ would be zero if there
were no entrainment, but in view of the effect originally predicted
by Andreev and Bashkin\cite{AndreevBashkin76} can be expected\cite{ALS84}
to be of the order of unity, while the effective London mass parameter
$\muL$ will have a magnitude that is the same as that of the relevant
charge carriers, in this case protons, to within a factor comparable with
unity, from which it differs by an amount that also depends on the
entrainment effect. The third parameter $\lL$ is interpretable as the 
relevant London penetration lengthscale that characterises the effective 
thickness of the individual proton vortices of the Abrikosov lattice. If 
it is assumed, as most authors have done,  that the entrainment effect 
only couples the neutrons and protons but does not significantly involve 
the leptonic background (so that the a master function  $\Lambda_{\rm M}$
of semi separable form\cite{CL98} can be used) then it can be estimated 
that this length scale will be given roughly as a function of the 
effective London mass $\muL$ and the lepton number density $\ne$ (which 
must be very close to the proton number density) by
 \beq \lL^2\simeq {\muL\over 4\pi e^2\ne}\, .\eqn{209}\eeq

It follows from (\ref{210}) that there will be a corresponding decomposition
 \beq \H_{\rho\sigma}=\H^{\rm n}_{\,\rho\sigma}+\HL_{\,\rho\sigma}
 +\HM_{\,\rho\sigma}\, ,\eqn{211}\eeq
with 
 \beq \H^{\rm n}_{\,\rho\sigma}= 2\nabla_{\![\rho}\A^{\rm n}_{\,\sigma]}
 \, , \hskip 1 cm \HL_{\,\rho\sigma} =2\nabla_{\![\rho}\AL_{\,\sigma]}
 \, , \hskip 1 cm \HM_{\,\rho\sigma}=2\nabla_{\![\rho}\AM_{\,\sigma]}
 \, , \eqn{212}\eeq
in which far as  the averaged flux is concerned, the main contribution
will typically be that of the London field $\HL_{\,\rho\sigma}$, meaning
the part attributable to the rotation of the ``normal'' (i.e. 
non-superfluid) negatively charged background. (In an ordinary metallic 
superconductor the analogous London field contribution arises as a well 
known consequence of rotation of the positively charged ionic background). 
In the absence of entrainment, the neutron vortex contribution 
$\H^{\rm n}_{\,\rho\sigma}$ would vanish. However the 
expectation\cite{ALS84} that the entrainment coefficient $\alpn$ will 
actually be of the order order unity implies that although it can be 
expected to be extremely small outside the immediate neighbourhood 
of a neutron vortex core (with a confinement radius of the same 
microscopic order of magnitude $\lL$ as that of a proton vortex) 
the integrated flux arising from neutron vortex contribution 
$\H^{\rm n}_{\,\rho\sigma}$ can be expected to be comparable with that
provided by the more smoothly distributed (unconfined) London contribution
 $\HL_{\,\rho\sigma}$. 

In normal circumstances, the least important term in  the sum (\ref{211}) 
will be what we shall refer to as the Meissner residue, meaning the
residual contribution $\HM_{\,\rho\sigma}$ arising from the 
semi-macroscopic current $\j^\rho$ if any. On the assumption that the
lengthscales characterising variation of the coefficients $\muL$ 
and $\lL$ are very long compared with the London penetration 
lengthscale $\lL$ itself, it can be seen that -- except within the 
microscopic defects forming the actual vortex cores where the mesoscopic 
superfluid description breaks down so that $\wn_{\rho\sigma}$ is locally 
non zero -- the dominant contribution in the source equation (\ref{86}) 
will be the part arising from the semi-macroscopic current $\j^\mu$ 
itself, so that, as a very good approximation, the source equation will 
reduce to the well known London form
 \beq\nabla^\sigma\nabla_{\,\sigma}\j^\rho\simeq {1\over\lL^2}\j^\mu
 \, ,\eqn{250}\eeq
whose homogeneous linear character entails that the only spacially and 
temporally uniform solution is that for which the current $\j^\mu$ simply
vanishes. What this implies is that after any initial high frequency 
osillations that may have been present have had time to radiate away 
or otherwise be dissipated, the medium will tend to settle towards 
a state in which the current actually is zero outside a very small
radius of order $\lL$ surrounding each individual vortex, so that its 
average $\langle\j^\rho\rangle$ over scales large compared with $\lL$ 
will also tend to vanish,
\beq \langle\j^\rho\rangle\simeq 0\, .\eqn{251} \eeq
The same  conclusion therefore applies to the corresponding
residual Meissner field contribution, $\HM_{\,\rho\sigma}$,
which will end up in a state such that
\beq \langle \HM_{\,\rho\sigma}\rangle\simeq 0\, .\eqn{252}\eeq

This last result is interpretable as a generalisation to ``type II''
(London - Abrikosov) superconductors of the Meissner effect
that was originally observed in laboratory examples of ``type I'',
meaning cases in which the Ginzburg Landau ratio of the penetration 
lengthscale $\lL$ to the relevant Pippard correlation lenghthscale 
is so small that, instead of condensing into an Abrikosov vortex lattice, 
the magnetic flux tends to be entirely expelled into domains where the 
superconductivity breaks down. In a type I situation the polarisation 
${\M}_{\rho\sigma}$ associated with the Abrikosov lattice will 
be absent, so there will be no distinction between the total flux 
$F_{\rho\sigma}$ and the contribution ${\H}_{\rho\sigma}$ as 
defined here, which means that in this experimentally more familiar case, 
expulsion of ${\H}_{\mu\nu}$ is equivalent to complete expulsion of 
$F_{\rho\sigma}$. On the other hand in the type II case, although there 
will be the same tendency to expulsion of $\H_{\rho\sigma}$, this will 
not entail the complete expulsion of $F_{\rho\sigma}$ because the
Abrikosov polarisation contribution ${\cal M}_{\mu\nu}$ will still
remain.

Both in the type II and -- as originally remarked by London -- in the 
type I case, the tendency for $\H_{\mu\nu}$ to vanish will be partially 
thwarted in a rotating background, for which the London contribution
$\HL_{\rho\sigma}$ will still remain even after the residual Meissner 
contribution $\HM_{\rho\sigma}$ has been dissipated in accordance
with (\ref{252}). In the usual laboratory applications, whether of type
I or type II, this London contribution is all that will remain, but in 
the neutron star case there will also be the neutron vortex contribution 
$\H^{\rm n}_{\rho\sigma}$. If the gradient of the (weakly density 
dependant) coefficient $\alpn$ is not entirely negligible, then as well 
as having a dominant part that is of magnetic character, may also 
include a small contribution to the electric displacement vector 
$\D_\rho$ as defined with respect to the ``normal'' leptonic background 
frame by the decomposition
 \beq \H_{\rho\sigma}=H_{\rho\sigma}+2u_{[\rho}\D_{\sigma]}\,  ,
 \hskip 1 cm\D_\rho=\H_{\rho\sigma}u^\sigma\, .\eqn{158}\eeq
Since the alignement of the neutron momentum covector will not on average
be very different from that of the background frame, it can be seen that 
what is to be expected is that under typical equilibrium conditions the 
macroscopically averaged value of the first contribution in (\ref{211}) 
will be given by
 \beq \langle\H^{\rm n}_{\,\rho\sigma}\rangle \simeq 2u_{[\rho}
 \langle\D^{\rm n}_{\sigma]}\rangle +\langle H^{\rm n}_{\rho\sigma}\rangle
 \, , \eqn{159}\eeq
in which the main contribution is the purely magnetic part given by
\beq H^{\rm n}_{\rho\sigma}\simeq
{1\over e}\langle\alpn \wn_{\rho\sigma} \rangle\, ,\eqn{155}\eeq
while the -- typically very small -- electric part will be given by
\beq \langle \D^{\rm n}_{\,\rho}\rangle \simeq
-{1\over e}\langle\mun\nabla_{\!\rho}\alpn\rangle \, ,\eqn{156}\eeq 
where the relevant neutron Fermi energy parameter is given
by $\mun= -u^\sigma \mun_\sigma$.

The analogous macroscopic average of the London contribution can be
instructively formulated in terms of the ``normal'' background's
acceleration tensor $\dot u^\rho$ and rotation tensor $\omeg_{\rho\sigma}$
(whose magnitude $\omeg=\sqrt{\omeg_{\rho\sigma}\omeg^{\rho\sigma}/2}$
is the local angular velocity) as defined by
 \beq \nabl_{[\rho}\,u_{\sigma]}=\omeg_{\rho\sigma}- u_{[\rho} 
 \dot u_{\sigma]}\, , \hskip 1 cm \dot u^\rho= u^\sigma\nabl_\sigma\, 
 u^\rho\, . \eqn{152}\eeq
The ensuing result (correcting a misplaced factor of 2 in the preceeding
presentation\cite{CL98}) is expressible as
 \beq \langle \H^{_{\rm L}}_{\rho\sigma}\rangle \simeq 2 u_{[\rho}\langle
 \DL_{\sigma]}\rangle+\langle H^{_{\rm L}}_{\rho\sigma}\rangle
 \, ,  \eqn{154}\eeq
in which the main contribution is the magnetic part which can be seen
to be given by
 \beq \langle H^{_{\rm L}}_{\rho\sigma}\rangle=-{2\over e}\langle\muL
 \omeg_{\rho\sigma}\rangle \eqn{153}\eeq
As originally observed by London,  this is proportional to the angular 
velocity $\omeg$, with a proportionality factor $\muL$ that in the 
neutron star application will be given roughly (but due to the 
``entrainment'' effect not exactly) by the proton mass (whereas in the 
ordinary metallic superconductors originally envisaged by London it 
is given by the electron mass). As well as this well known magnetic 
contribution, there will also be a corresponding, but typically much 
less important, electric displacement contribution given by
 \beq \langle \DL_\rho\rangle \simeq {1\over e}\langle
 \muL\dot u_\sigma+ \nabla_{\!\sigma}\muL\rangle\, .\eqn{157}\eeq
(This electric displacement field is usually ignored in discussions of 
laboratory applications, but even in the ideally simplified case of a 
motionless incompressible sample that is postulated to be strictly 
homogeneous so that the gradient term would be absent, a small residual 
electric field of this type would still be needed to balance the 
effect on the conducting particles -- which in that case would be 
electrons -- of the ordinary terrestrial gravitation field.)

I wish to thank Silvano Bonazzola, David Langlois, and Reinhard Prix for 
many helpful discussions, and I particularly wish to express
my appreciation to Isaac Khalatnikov for instructive 
conversations on numerous occasions since my interest in this subject was 
originally inspired by his classic textbook\cite{Kha65}.
 
\vfill\eject


\begin{thebibliography}{99}

\bibitem{C85} B. Carter,
``The canonical treatment of heat conduction and superfluidity in 
relativistic hydrodynamics'',
in {\it A random walk in Relativity and Cosmology (Essays in honour of P.C.
Vaidya \& A.K. Raychaudhuri)}, ed. N. Dadhich, J. Krishna Rao, J.V. Narlikar,
C.V. Visveshwara, {\it pp} 48-62 (Wiley Eastern, Bombay, 1985).

\bibitem{CL98} B. Carter, D. Langlois,
``Relativistic model for superconducting superfluid mixtures'',
{\it Nucl. Phys.}  {\bf B531}, {\it pp} 478-504 (1998)
[gr-qc/9806024].

\bibitem{LK82} V.V. Lebedev, I.M. Khalatnikov,
``Relativistic hydrodynamics of a superfluid'', 
{\it Sov. Phys. J.E.T.P.} {\bf 56}, {\it pp} 923-930 (1982).

\bibitem{CK92b} B. Carter, I.M. Khalatnikov,
``Momentum, Vorticity, and Helicity in Covariant Superfluid Dynamics",
{\it Ann. Phys.} {\bf 219}, {\it pp} 243-265 (1992).

\bibitem{CL95c} B. Carter, D. Langlois, 
``Kalb-Ramond coupled vortex fibration model for relativistic superfluid
dynamics'', 
{\it Nuclear Physics} {\bf B 454}, 402-424 (1995)
[hep-th/9611082].

\bibitem{AndreevBashkin76} A.F. Andreev, E.P. Bashkin,
``Three velocity hydrodynamics of superfluid solutions''
{\it Sov. Phys. J.E.T.P.,} {\bf 42}, 164-646 (1975). 

\bibitem{VarSed81}G.A. Vardanyan, D.M. Sedrakyan,
``Magnetohydrodynamics of superfluid solutions'',
{\it Sov. Phys. J.E.T.P.} {\bf 54}, 919-921 (1981).

\bibitem{MenLin91} G. Mendell, L. Lindblom,
``The coupling of charged superfluid mixtures to the electromagnetic field'',
{\it Ann. Phys.} {\bf 205}, 110-129 (1991).

\bibitem{BekaK61} I.L. Bekarevich and I.M. Khalatnikov,
``Phenomenological derivation of the equations of vortex motion in HeII'',
{\it Sov. Phys. J.E.T.P.} {\bf 13}, 643 (1961).

\bibitem{CPL99} B. Carter, R. Prix, D. Langlois,
``Energy of Magnetic Vortices in Rotating Superconductor''
{\it Phys. Rev.} {\bf B62} (2000) [cond-mat/9910240]. 

\bibitem{ALS84} M.A. Alpar, S.A. Langer, J.A. Sauls,
``Rapid postglitch spin-up of the superfluid core in pulsars'',
{\it Astroph. J.} {\bf 282}, 533-541 (1984).

\bibitem{CPL99b} B. Carter, D. Langlois, R. Prix,
``Bogomol'nyi limit for magnetic vortices in rotating superconductor'',
{\it Phys. Rev.} {\bf B62} (2000) [cond-mat/9910263].

\bibitem{Men91} G. Mendel,
``Superfluid hydrodynamics in rotating neutron stars. I Nondissipative
equations'', {\it Astroph. J} {\bf 380}, {\it pp} 515-529 (1991).

\bibitem{Men98} G. Mendel,
``Magnetohydrodynamics in superconducting-superfluid neutron stars'',
{\it Mon. Not. R. Astron. Soc.} {\bf 296}, {\it pp} 903-912 (1998)
[astro-ph/9702032].

\bibitem{Kha65} I. M. Khalatnikov,
{\it Introduction to the Theory of Superfluidity}
(Benjamin, New York, 1965).

\end{thebibliography}
\end{document}